\documentstyle[12pt]{article}
\begin{document}

\title{A singularity-free space-time\vspace{.3cm}}

\author{F.J.Chinea and L.Fern\'andez-Jambrina\vspace{.3cm}\\
   Departamento\ de F\'{\i}sica Te\'orica II,\\
Facultad de Ciencias F\'{\i}sicas,\\
 Universidad Complutense, 28040 Madrid, Spain\\
and\\
Instituto\ de F\'{\i}sica Fundamental,\\
Ciudad Universitaria, 28040-Madrid, Spain\vspace{.5cm}
\and
J.M.M.Senovilla\vspace{.3cm}\\
Departament de F\'{\i}sica Fonamental\\
Universidad de Barcelona\\ Diagonal 647, 08028 Barcelona,
Spain\\
and\\
Laboratori\ de F\'{\i}sica Matem\`atica,\\
Societat Catalana de F\'{\i}sica, I.E.C., Barcelona,
Spain}
\date{}
 \maketitle
\begin{abstract}
We show that the solution published in Ref.1 is
geodesically complete and singularity-free. We also prove
that the solution satisfies the stronger energy and
causality conditions, such as global hyperbolicity,
causal symmetry and causal stability. A detailed
discussion about which assumptions in the singularity
theorems are not fulfilled is performed, and we show
explicitly tat the solution is in accordance with those
theorems. A brief discussion of the results is given.

\end{abstract}
\hfil

\newpage

\section{Introduction}
Recently, one of us presented a solution to Einstein's field equations for a
perfect-fluid energy-momentum tensor without any curvature singularity (Ref.1).
The solution has positive pressure and energy density everywhere, and it
satisfies
a realistic equation of state for hot radiation-dominated epochs. As we shall
see,
it also verifies the stronger causality requirements. Therefore, the question
remains
of whether or not the solution is geodesically complete and, if it is, how that
fits in with the general conclusions of the very powerful singularity theorems
[2,3].

The main result in this paper is that the mentioned solution is geodesically
complete
and, in fact, singularity-free (see Ref.2 for definitions). We shall prove
it in
Sec.2 by studying the equations of the non-spacelike geodesics and showing that
they can always be extended to arbitrary values of the affine parameter. Once
this important property has been established, in Sec.3 we shall study which of
the assumptions
of the different singularity theorems are not verified in the solution. Since
the
energy and causality conditions will be proven to be satisfied, it will turn
out
that the so-called initial or boundary condition is the one that fails to hold.
This condition usually refers to the existence of a causally trapped set [2,3]
or to some fixed bound for the initial expansion of the geodesics. Thus, we shall
see that the initial condition is {\it essential\/} for the development of
singularities. Along with this analysis, we shall prove a number of nice and
desirable properties of the solution, among them, global hyperbolicity and stable
causality. Finally,
we devote Sec.4 to give a brief discussion of the results herein presented.

Before getting into the main subject, let us give a summary of the solution
of [1]. The line element is given by
\begin{eqnarray*}
ds^2 = 
C^4(at)C^2(3ar)(-dt^2+dr^2)+(9a^2)^{-1}C^4(at)S^2(3ar)C^{-2/3}(3ar)d\phi^{2}+
\end{eqnarray*}     \vspace{-1 cm}
\begin{eqnarray} \hspace{-5 cm}
 C^{-2}(at)C^{-2/3}(3ar)dz^2, \end{eqnarray}
where $a\neq 0$ is an arbitrary constant (which can be taken as positive) and we use
the notation \begin{eqnarray*}
S(u)\equiv sinh(u),\hspace{0.5 cm}  C(u)\equiv cosh(u).
\end{eqnarray*}
The coordinates used are of cylindrical type, and their range is
\begin{eqnarray*}
-\infty <t,z<\infty ,\hspace{0.5 cm}  0\leq r<\infty , \hspace{0.5 cm} 0\leq \phi
\leq 2\pi \end{eqnarray*}
such that $\phi $ and $\phi +2\pi $ are identified. The metric (1) is a solution
of Einstein's equations for a perfect fluid. The coordinates are comoving (or
adapted to the fluid congruence) in the sense that the velocity vector has the
form \begin{equation} {\bf u}=C^{-2}(at)C^{-1}(3ar)\partial /\partial t.
\end{equation}
The fluid congruence is orthogonal to the spacelike hypersurfaces $t=const.$ and
therefore its rotation tensor vanishes. The expansion of the fluid has the expression
\begin{equation}
\theta =3aS(at)C^{-3}(at)C^{-1}(3ar),
\end{equation}
which is regular everywhere and positive for $t>0$ (expanding phase) or negative
for $t<0$ (contracting phase). We also see that both for $t\rightarrow \pm \infty$
or $r\rightarrow \infty $ the expansion approaches zero.

However, the congruence of the fluid is not geodesic, the acceleration vector field
 being \begin{equation} {\bf a}=3aS(3ar)C^{-3}(3ar)C^{-4}(at)\partial /\partial r.
\end{equation}
From a physical point of view, this fact is of extreme importance for the absence
of singularities in the solution. Because, as is well known, a non-vanishing
acceleration means that there exists a spacelike {\it gradient of pressure\/} or
equivalently, a force which opposes to the gravitational attraction. Thus, although
for $t<0$ there is a contraction, the gradient of pressure does not allow for the
formation of a singularity and causes a bounce from which the matter expands.
The fluid congruence is also shearing, the shear tensor can be found in [1].

The density and pressure of the fluid take the following form
\begin{eqnarray}
\chi \rho =15a^2C^{-4}(at)C^{-4}(3ar),\\
p=\rho /3
\end{eqnarray}
where $\chi $ is the gravitational constant. From (5) we see that the maximum
value of $\rho $ at each time $t$ is achieved at $r=0$. Moreover, the absolute
maximum of the density (and pressure) occurs when $r=0$ and $t=0$, and its value
is $\rho _M=15a^2/\chi $. Therefore, we can interpret the constant $a$ as the
maximum energy density. Note that, since $a$ is arbitrary, the maximum density can be
chosen as large or small as we like. On the other hand, from (6) we learn that
the equation of state is realistic for a space-time filled with incoherent radiation
Equations (5) and (6) also tell us that $\rho $ and $p$ are positive everywhere,
which implies that the strong energy condition (or the timelike convergence
condition, see [2]) is satisfied. From the same equations we can check that $\rho $
and $p$ are finite and well-behaved everywhere, which means that there is no matter
singularity (singularity in the Ricci tensor). It can further be shown, by compu
ting the Weyl tensor (see Ref.1), that {\it all\/} the curvature invariants are
regular on the whole space-time so that there is no curvature singularity in the
metric at all.

The line element (1) admits an abelian group of symmetries $G_2$, and the Killing
vectors $\partial /\partial \phi $ and $\partial /\partial z$ are globally
defined. Both of them are spacelike, orthogonal to each other and orthogonally
transitive, and therefore the metric is cylindrically symmetric. The regularity
condition [4] on the axis $r=0$ is satisfied so that the $2\pi $-periodicity of the
coordinate $\phi $ is well defined and the so-called elementary flatness on the
vicinity of the axis is assured. In this sense, the coordinate singularity
which appears in (1) at $r=0$ is just that of cylindrical coordinates; other
coordinates (Cartesian-like) do exist such that they provide a global chart
on the manifold.

The coordinate $t$ appearing in the metric is a {\it time function\/} (or cosmic
time) in the sense that it (decreases) increases along every (past-)
future-directed non-spacelike curve. This can be seen by checking that the gradient
of $t$ is timelike everywhere. As is well known, this is the necessary and
sufficient condition for the stable causality condition to hold [2]. Thus, the
space-time is causally stable which is the stronger causality requirement and it
implies the weaker chronology and causality conditions.

Finally, let us remark that the solution is inextendible, which avoids the
possibility of the existence of singularities hidding somewhere. This is
apparent from the form of the metric (1); however, the best way to prove it is
to show that all geodesics are complete and then the metric can be extended nowhere.
In the next section we shall prove geodesic completeness for the solution.
\section{Geodesic completeness}
Our aim now is to show that every non-spacelike geodesic can be extended to
arbitrary values of its affine parameter. To that end, we must analyze the equations
of the geodesics in the metric (1), which after standard and straightforward
calculations can be written as follows
\begin{eqnarray}
\ddot{t}+2aT(at)(\dot{t}^2+\dot{r}^2)+6aT(3ar)\dot{t}\dot{r}+ \nonumber \\
(2/9a)T(at)S^2(3ar)C^{-8/3}(3ar)\dot{\phi }^2- \nonumber \\
aC^{-7}(at)S(at)C^{-8/3}(3ar)\dot{z}^2=0,
\end{eqnarray}
\begin{eqnarray}
\ddot{r}+3aT(3ar)(\dot{t}^2+\dot{r}^2)+4aT(at)\dot{t}\dot{r}- \nonumber \\
(1/9a)S(3ar)C^{-5/3}(3ar)[3-T^2(3ar)]\dot{\phi }^2+ \nonumber \\
aC^{-6}(at)S(3ar)C^{-11/3}(3ar)\dot{z}^2=0,
\end{eqnarray}
\begin{eqnarray}
(9a^2)^{-1}C^4(at)S^2(3ar)C^{-2/3}(3ar)\dot{\phi }=K, \\
C^{-2}(at)C^{-2/3}(3ar)\dot{z}=L,
\end{eqnarray}
\begin{eqnarray}
C^4(at)C^2(3ar)(-\dot{t}^2+\dot{r}^2)+ \nonumber \\
(9a^2)^{-1}C^4(at)S^2(3ar)C^{-2/3}(3ar)\dot{\phi }^2+ \nonumber \\
C^{-2}(at)C^{-2/3}(3ar)\dot{z}^2=-\delta .
\end{eqnarray}
Here a dot means derivative with respect to the affine parameter and we use the
notation
\begin{eqnarray*}
T(u)\equiv tanh(u).
\end{eqnarray*}
$K$ and $L$ are constants of motion along the geodesics due to the existence of
two Killing vectors in the spacetime. Another constant of motion is denoted by
$\delta $, which must be taken equal to one or zero for timelike or null geodesics
respectively. The fact that all functions involved in eqs.(7)-(11) are
non singular will grant us the existence and uniqueness of solutions.

The method of the proof is to give finite bounds for the first derivatives of
the coordinates, which implies (see Ref.5) that the field is non-singular and
the geodesics are complete. We will also use the second derivatives of the
coordinates to show that these cannot become singular. We shall only deal with
geodesics propagating towards the future. Propagation towards the past can be
treated in a similar way.

In equation (8), when (9) and (10) are taken into account,
$\ddot r$ is negative for positive $t$ and increasing values of $r$. This follows
from the fact that the positive term in $\dot z^2$ dominates over the negative
term in $\dot \phi ^2$ for large enough values of $r$. Negative values of $t$
need not be considered, as $t$ will end up becoming positive. It is obvious that $r$
cannot diverge to infinity in a finite proper time with negative $\ddot r$.

Most of the geodesics approach $r=0$ without any problem. As $r$ decreases, the
term in $\dot t \dot r$ dominates over the terms in $\dot r^2$ and $\dot t^2$,
while the term in $\dot z^2$ vanishes when approaching $r=0$. Hence, $\ddot r$ is
positive for decreasing $r$ in the vicinity of $r=0$ for positive $t$. This means
that the geodesics cannot come near the axis quickly enough to become singular.

Also $\ddot t$ is negative for large values of $t$ (notice that the term in $\dot t
\dot r$ would not be relevant if $r$ decreases), so that the geodesic does not grow
faster than its tangent. The quantity $\dot z$ does not become singular if $r$ and
$t$ do not; $\dot \phi$ could have problems if $r$ were zero, but we shall see that
this is not possible.
 In order to
proceed, we divide our study in steps, starting with the simpler geodesics and then
going to the general case. In that way we shall be able to use some results several
times as we progress on.

1) {\it Geodesics in the fluid congruence\/}. This is the simplest case. The
congruence of the fluid is defined by $\dot{r}=\dot{\phi }=\dot{z}=0$. From
Eq.(4) it follows that the only possible geodesics lie in $r=0$. Then the system
reduces to the single equation
\begin{eqnarray*}
\dot{t}=C^{-2}(at)\leq 1.
\end{eqnarray*}
Obviously, these geodesics are complete.

2) {\it Geodesics along the axis\/}. They are defined by $\dot{r}=\dot{\phi }=0$
$r=0$. The remaining equations are now
\begin{eqnarray*}
\dot{z}=LC^2(at), \\
\dot{t}=C^{-2}(at)[L^2C^2(at)+\delta ]^{1/2}\leq (L^2+\delta )^{1/2}.
\end{eqnarray*}
Therefore, the derivatives are bounded everywhere and these geodesics are also
complete. The coordinate $t$ reaches infinity only when so does the affine
parameter. This is a general property of all geodesics and we shall not refer
to it in what follows.

3) {\it Radial null geodesics\/}. Now $\dot{\phi }=\dot{z}=0$, $\delta =0$. After
one integration the system reduces to
\begin{eqnarray*}
\dot{t}=\vert \dot{r}\vert , \hspace{0.5 cm} \dot{r}=hC^{-4}(at)C^{-2}(3ar)
\end{eqnarray*}
where $h$ is a constant of integration. We see that
$0<\vert \dot{r}\vert =\vert \dot{t}\vert \leq \vert h\vert $, from where it follows
that these geodesics are complete too. When one of the curves arrives at $r=0$ along
a direction $\phi $ then it continues along the direction $\phi +\pi $ after
crossing the axis without problems.

4) {\it Radial timelike geodesics\/}. Again $\dot{\phi }=\dot{z}=0$, but now
$\delta = 1$. By introducing a new function $v$ which, so to speak, parametrizes
$\dot{t}$ and $\dot{r}$, the system reduces in this case to
\begin{eqnarray}
\dot{t}=C(v)A(r,t),\hspace{0.3 cm} \dot{r}=S(v)A(r,t), \hspace{0.3 cm}
\dot{v}=B(r,t,v)A(r,t) \end{eqnarray} where
\begin{eqnarray}
A(r,t)\equiv C^{-2}(at)C^{-1}(3ar), \\
B(r,t,v)\equiv -a[3T(3ar)C(v)+2T(at)S(v)].
\end{eqnarray}
The previous reasoning about second derivatives allows us to conclude that these
geodesics are complete.

5) {\it Null geodesics with no angular velocity\/}. Now,
$\dot{\phi }=0$ and $\delta =0$. Using the function $v$ again, we can write the
system of equations \begin{eqnarray*}
\dot{t}=C(v)E(r,t),\hspace{0.3 cm} \dot{r}=S(v)E(r,t),\hspace{0.3 cm}
\dot{v}=E(r,t)F(r,t,v) \end{eqnarray*}
where
\begin{eqnarray}
E(r,t)\equiv \vert L\vert C^{-1}(at)C^{-2/3}(3ar), \\
F(r,t,v)\equiv -a[4T(3ar)C(v)+3T(at)S(v)].
\end{eqnarray}
A reasoning similar to that given in the previous case leads us again to
completeness for these null geodesics.

6) {\it Null geodesics on the hypersurfaces z=const.}. These are defined by
$\dot{z}=0$, $\delta =0$. The system of equations can be written as
\begin{eqnarray*}
\dot{t}=C(v)M(r,t),\hspace{0.3 cm} \dot{r}=S(v)M(r,t),\hspace{0.3 cm}
\dot{v}=M(r,t)D(r,v) \end{eqnarray*}
with
\begin{eqnarray}
M(r,t)\equiv 3\vert aK\vert C^{-4}(at)C^{-2/3}(3ar)S^{-1}(3ar), \\
D(r,v)\equiv aC(v)[3T^{-1}(3ar)-4T(3ar)].
\end{eqnarray}
In this case we can obtain one of the orbit equations by dividing $\dot{r}$ by
$\dot{v}$ and then integrating. Doing so we get
\begin{eqnarray*}
C(v)=\alpha ^{-1}S(3ar)C^{-4/3}(3ar)
\end{eqnarray*}
where $\alpha $ is a positive constant. We see that, since $C(v)\geq 1$, the
coordinate $r$ can only take values between $r_+$ and $r_-$, with $r_{\pm }$
defined by $S(3ar_{\pm })C^{-4/3}(3ar_{\pm })=\alpha $. Thus $\vert v\vert $
is bounded and the geodesics are complete.

Among these geodesics there is one which is circular and goes round the axis of
symmetry at $r=r_o\equiv arccosh(2)/3a$. This can be identified as the special
case in which $r_+=r_-=r_o$, $v=0$.

7) {\it General non-spacelike geodesics\/}. Using the $v$-parametrization, we
can rewrite the general system of equations (7)-(11) in the following form
\begin{eqnarray*}
\dot{t}=(3\vert aK\vert )^{-1}M(r,t)G(r,t)C(v),\hspace{0.3 cm} \dot{r}=(3\vert
aK\vert)^{-1}M(r,t)G(r,t)S(v), \end{eqnarray*}
\begin{eqnarray}
\dot{v}=(3\vert aK\vert )^{-1}M(r,t)H(r,t)G^{-1}(r,t)
\end{eqnarray}
where we have defined
\begin{eqnarray*}
G(r,t)\equiv [9a^2K^2+L^2C^6(at)S^2(3ar)+C^4(at)S^2(3ar)C^{-2/3}(3ar)(1-\delta
)]^{1/2}, \end{eqnarray*}
\begin{eqnarray}
\end{eqnarray}
\begin{eqnarray}
H(r,t,v)\equiv -a\{ S(v)T(at)S^2(3ar)[3L^2C^6(at)+2C^4(at)C^{-2/3}(3ar)(1-\delta
)]+ \nonumber \\
 C(v)T(3ar)[4L^2C^6(at)S^2(3ar)+3C^4(at)S^2(3ar)C^{-2/3}(3ar)(1-\delta
)]+ \nonumber \\ 9a^2K^2C(v)[4T(3ar)-3T^{-1}(3ar)]\}.\nonumber \\
\end{eqnarray}
The only problem we could have appears when $r$ approaches zero, but in this case
the dominant terms are those of paragraph 6), so that the geodesics with $K \neq 0$
cannot reach the axis. Then they are complete. We can say that the term in $\dot
\phi ^2$ in (8) is centrifugal in the vicinity of the axis.
 \section{Properties of
the solution and its relation with singularity theorems}
 From
the results of the previous section it is obvious that {\it every\/} maximally
extended null geodesic meets any of the hypersurfaces $t=const.$. This is a
sufficient condition such that every non-spacelike curve intersects the mentioned
hypersurfaces exactly once [6]; in other words, the hypersurfaces $t=const.$ are
global Cauchy surfaces and the solution is globally hyperbolic. For this reason the
solution is also causally simple [2], that is, for every compact set $K$, $J^+(K)$
is closed and its boundary coincides with $E^+(K)$.

Due to the fact  that every spacelike hypersurface $t=const.$ is a global
Cauchy surface, and since $t$ is a time function in the solution as explained
in Sec.1, it follows that {\it every\/} non-spacelike curve (geodesic or not)
can be extended to arbitrary values of its generalized affine parameter (because
it has to meet {\it all\/} the Cauchy surfaces). This means that the solution
is non-spacelike $b$-complete (see Ref.2), that is to say, {\it singularity-free}.

Furthermore, from expressions (5) and (6) we see that the energy-momentum tensor
does not vanish anywhere. Therefore, given any non-spacelike vector {\bf $v$}
we have $R_{ab}v^av^b>0$ so that, in addition to the strong energy condition, the
generic condition is fulfilled as well [3]. Thus, it remains to see which other
conditions in the singularity theorems are not satisfied by the spacetime.

The simplest such theorem (Ref.2, p.274) requires the existence of a Cauchy surface
such that the trace of its second fundamental form is bounded away from zero.
In our case, the mentioned trace for any Cauchy surface $t=const.$ is given by
the expansion of the fluid congruence shown in (3). From this expression is
obvious that $\theta $ is not bounded away from zero, because, for any $t$,
$\theta $ goes to zero when $r\rightarrow \infty $. The significance of this
result can be better understood as follows: given any Cauchy surface $t=t_1<0$, we
have $\theta _{t_1}<0$ and then there are points conjugate to the surface along
every future-directed timelike geodesic orthogonal to the surface within a distance
$-3/\theta _{t_1}$. However, as $\theta _{t_1}$ is not bounded away from zero, those
distances do not have an upper bound. This allows for the existence of a maximal
geodesic orthogonal to the surface up to any point $q$ to its future, as is necessary
being $t=t_1$ a Cauchy surface.

Another property of the solution is that the Cauchy surface $\Sigma :t=0$ is a
maximal spacelike hypersurface, i.e., one in which the trace of the second
fundamental form vanishes. It follows that the spacetime is causally (or time)
symmetric [7] so that $J^+(\Sigma )$, $I^+(\Sigma )$ and $D^+(\Sigma )$ are
isometric to $J^-(\Sigma )$, $I^-(\Sigma )$ and $D^-(\Sigma )$ respectively.
In Ref.7, Tipler has shown that all geodesics are both future and past incomplete
in a spacetime which contains a non-compact maximal Cauchy surface if, in addition
to the strong energy condition, the following is satisfied: there exist fixed
positive constants $b,c$ such that
\begin{eqnarray*}
\vert \int _o^b R_{ad}v^av^d d\tau \vert \geq c
\end{eqnarray*}
for every timelike geodesic (with tangent vector {\bf $v$}) intersecting $\Sigma$
orthogonally at $\tau =0$, where $\tau $ is the affine parameter. However, this
is not verified in solution (1), because given any pair of constants $b$ and $c$,
the above integral is always positive but not bounded below above zero. In fact,
we can always choose geodesics (for $r$ big enough initially) such that the
integral takes a value as small as we like and less than any previously fixed $c$.

Let us consider then the classical singularity theorems. The first of them,
chronologically speaking, is that of Penrose [8,2], which assumes the existence
of both a non-compact Cauchy surface and a closed trapped surface, i.e. a compact
(without boundary) spacelike two-surface in which the traces of the two null
second fundamental forms have the same sign. Our space-time does not have a closed
trapped surface. In order to prove it, suppose there was one. Since the surface is
compact, it must have a point $p$ where $r$ reaches its maximum, so that the normal
at $p$ is a superposition of ${\partial  \mathord{\left/ {\vphantom {\partial 
{\partial r}}} \right. \kern-\nulldelimiterspace} {\partial r}}$ and
${\partial  \mathord{\left/ {\vphantom {\partial 
\partial }} \right. \kern-\nulldelimiterspace} \partial }t$ at $p$. But if we
compute the traces of both null second fundamental forms at $p$, we get

 \begin{eqnarray*}
\chi _{\mp \;
d}^d=-g^{zz}n_{z,z}(p)-g^{\phi
\phi}n_{\phi ,\phi}(p)+\\2^{-1/2}aC^{-2}(at)S^{-1}(3ar)[-T(at)T(3ar)\mp 3\pm
2T^2(3ar)] \end{eqnarray*} from where it is clear that \begin{eqnarray*} \chi _{-\;
d}^d\leq -2^{-1/2}aC^{-2}(at)S^{-1}(3ar)[1+T(at)T(3ar)]<0,
\\ \chi _{+\; d}^d\geq
2^{-1/2}aC^{-2}(at)S^{-1}(3ar)[1-T(at)T(3ar)]>0,
\end{eqnarray*} since $n_{z,z}(p)$ and $n_{\phi ,\phi}(p)$
are positive for outgoing normals and negative for ingoing
normals at $p$. We see that the traces have opposite signs
so that there are no closed trapped surfaces. This
reasoning can be visualized in the following way. Take any
closed compact two-surface in the manifold. This surface
must be orthogonal to the in- and out-going radial null
geodesics somewhere. But the outgoing and ingoing radial
null geodesics are expanding and contracting, respectively,
everywhere.

The most famous and powerful singularity theorem was proven by Hawking and
Penrose [9,2]. Our space-time satisfies all conditions in the Hawking-Penrose
theorem except for condition 4 as appears in Ref.2, which allows for three
different possibilities. One of them is the existence of a closed trapped
surface, which we have just shown is not satisfied in the solution. The second
possibility is the existence of a point $q$ such that on {\it every\/} past
(future) null geodesic from $q$ the expansion becomes negative. There is no
such point in solution (1), since through any point in the manifold there are
radial null geodesics which diverge if they are outgoing and future-directed or
ingoing and past-directed. Another way to see that the point $q$ does not exist is
to remember that through any $q$ there are null geodesics with $z=const.$ which are
bounded above and below in $r$. Thus, these geodesics can never converge with the
radial ones. The third possibility in Hawking-Penrose's theorem is the existence of
a compact achronal set without edge. It is pretty obvious that there is not a set
with those properties in the space-time under consideration. To prove it rigourosly,
take any achronal set in the manifold and a point $q$ in the set. By using the
radial geodesics we can always choose points $q_-\in I^-(q)$ and $q_+\in I^+(q)$
such that $r(q_-)=r(q_+)>r(q)$, where for any point $s$ we denote by $r(s)$ the value
of the coordinate $r$ at $s$. Since $q_+\in I^+(q_-)$ and $r(q_+)=r(q_-)$, we can
join $q_-$ and $q_+$ with a future-directed worldline of the fluid congruence. If the
achronal set has no edge, this worldline {\it must\/} intersect the set, and it will
do it at a point $\tilde{q}$ with $r(\tilde{q})=r(q_-)=r(q_+)>r(q)$. We have thus
proven that for any achronal set without edge, the coordinate $r$ cannot be bounded.
It follows that any achronal set in the manifold cannot be both compact and
without edge.

Finally, let us remark that in Ref.2 there appear two other singularity theorems
due to Hawking (marked with numbers 3 and 4, pages 271 and 272). However, the
conditions in these theorems are stronger than those of the Hawking-Penrose theorem
which we have already proven not to hold in the solution. There are some other
singularity theorems now available, but their hypothesis are mere variations
of those here studied. In this sense, we believe that our study is somehow
exhaustive. In the next section we shall give a brief discussion of the meaning of
our results.
\section{Discussion}
We have proven that solution (1) is singularity-free and in accordance with the
main singularity theorems {\it as well\/}. In fact, once the properties of the
solution are known, it becomes rather obvious that it is free of singularities.
Because the solution is globally hyperbolic so that there cannot be any Cauchy
horizon. However, the congruence of the fluid is trivially complete and through
each point in the manifold there is a worldline of the fluid congruence.

It is also clear that, due to the properties of the solution, any possible
singularity would have to have some extension and therefore it should manifest
itself in the curvature invariants; but they do not. However, to be on the safe
side, we have preferred to perform a detailed analysis of which particular
condition in the different singularity theorems failed to be verified in the
space-time for the following reasons. First, to illustrate how the solution can
avoid the development of singularities. Second, to get a deeper insight in the
significance and application of the singularity theorems and, in this sense, to show
explicitly that neither the energy or causality conditions are determinant by
themselves for the appearance of singularities, not even for solutions filled with
reasonable matter everywhere and such that all the matter is expanding at a given
instant of time.

As is well known, reasonable singularity-free solutions are a rarity at the
moment. This is specially true for solutions with cosmological properties. By this
we do not mean solutions which can describe adequately the real observed
Universe (which is not the case here), but rather solutions which {\it
theoretically} must be considered as cosmological since they have matter everywhere
and cannot be considered as interiors of some vacuum exteriors (for this it
would be necessary a timelike surface of vanishing pressure). Therefore, the
question arises of how many singularity-free solutions there possibly are and
which particular properties they must obey.

It can be thought, for example, that the special equation of state (6) would
have some importance in order to avoid the singularities. This is not true, for in
Ref.10 it has been shown that the solution here studied belongs to a larger family of
singularity-free metrics for perfect fluids with no equation of state whatsoever.
However, all the members in the family have in common that they are cylindrically
symmetric and, in fact, every other solution found in [10] without this symmetry
contains singularities. Thus, cylindrical symmetry could be somehow important
for the avoidance of singularities. Of course, we must keep this conclusion as a mere
hypothesis. On the other hand, the converse of this hypothesis is obviously
false, and, in fact, if we replace $cosh(at)$ for $sinh(at)$ in (1) we obtain another
solution of Einstein's equations with similar properties and a well-defined
cylindrical symmetry but having a big-bang singularity in the finite past [1,11].

In any case, it is not clear to us why there should not be general solutions
(with no symmetry) without singularities if they fulfill the main properties
shown in this paper, that is to say, if they do not have any kind of causal
trapped set, which reveals itself as an essential assumption in the singularity
theorems.\\

\vspace{1.5 cm} \noindent
{\Large\bf Acknowledgments}\\
\vspace{0.5 cm}

\noindent
The present work has been supported in part by DGICYT Project PB89-0142 (F.J.C.)
and CICYT Project AEN90-0061 (J.M.M.S); L.F.J. is supported by a FPI Predoctoral
Scholarship from Ministerio de Educaci\'{o}n y Ciencia (Spain). J.M.M.S. wishes to
thank E. Ruiz for discussions.

\newpage
{\bf References}

[1] J.M.M. Senovilla, Phys. Rev. Lett. {\bf 64}, 2219 (1990).

[2] S.W. Hawking and G.F.R. Ellis, {\it The large scale structure

of space-time\/}, Cambridge Univ. Press, Cambridge (1973).

[3] J. Beem and P. Ehrlich, {\it Global Lorentzian Geometry\/},

Dekker, New York (1981).

[4] D. Kramer, H. Stephani, M. MacCallum and E. Herlt, {\it Exact

Solutions of Einstein's Field Equations\/}, Cambridge Univ. Press,

Cambridge (1980).

[5] V.I. Arnold, {\it Ordinary Differential Equations\/},

MIT Press (1973).

[6] R. Geroch, J. Math. Phys. {\bf 11}, 437 (1970).

[7] F.J. Tipler, J. Math. Phys. {\bf 18}, 1568 (1977).

[8] R. Penrose, Phys. Rev. Lett. {\bf 14}, 57 (1965).

[9] S.W. Hawking and R. Penrose, Proc. Roy. Soc. London A{\bf 314},

 529 (1970).

[10] E. Ruiz and J.M.M Senovilla, to be published.

[11] A. Feinstein and J.M.M. Senovilla, Class. Quantum Grav. {\bf 6},

 L89 (1989).

\end{document}